\documentclass[twocolumn,superscriptaddress,floatfix,prb]{revtex4}
\usepackage{graphicx,amsfonts,amssymb,amsmath,hyperref,hypcap,enumerate}
\usepackage[mathscr]{euscript}
\usepackage{color}

\newcommand{\kk}{\boldsymbol{k}}

\begin{document}
\title{Majorana Kramers pair in a nematic vortex}

\author{Fengcheng Wu}
\affiliation{Materials Science Division, Argonne National Laboratory, Argonne, IL 60439, USA}

\author{Ivar Martin}
\affiliation{Materials Science Division, Argonne National Laboratory, Argonne, IL 60439, USA}

\date{\today}

\begin{abstract}
A time-reversal (TR) invariant topological superconductor is characterized by a Kramers pair of Majorana zero-energy modes on boundaries and in a core of a TR invariant vortex.  A vortex defect that preserves TR symmetry has remained primarily of theoretical interest, since typically a magnetic field, which explicitly breaks TR, needs to be applied to create vortices in superconductors. In this work, we show that an odd-parity topological superconductor with a nematic pairing order parameter can host a {\em nematic vortex} that preserves TR symmetry and binds a Majorana Kramers pair. Such a nematic superconductor could be realized in metal-doped Bi$_2$Se$_3$, as suggested by recent experiments.  We provide an analytic solution for the zero modes in a continuous nematic vortex. In lattice, crystalline anisotropy can pin the two-component order parameter along high-symmetry directions. We  show that a discrete nematic vortex, which forms when three nematic domains meet, also supports a TR pair of Majorana modes.
Finally, we discuss possible experiments to probe the zero modes. 
\end{abstract}

\maketitle

\section{Introduction}

Topological superconductors represent a paradigmatic system where topological effects and many-body interactions can have an interesting interplay. Topological superconductors with broken time reversal (TR) symmetry support Majorana zero-energy modes localized at the boundaries\cite{kitaev2001} and in the vortex cores\cite{Read2000, Ivanov2001}.  The non-Abelian  statistics of Majorana modes may be utilized for fault-tolerant quantum computation\cite{kitaev2003}. The search for Majorana modes has been greatly influenced by the theoretical proposals based on topological insulator surface states\cite{Fu_Kane} or Rashba wires\cite{Sau2010,Lutchyn2010} in proximity to ordinary $s$-wave superconductors  in presence of magnetic field. Exciting experimental progress\cite{albrecht2016, Sun2016} has been made along the lines of these proposals. When TR symmetry is preserved,  the Kramers pairs of Majorana modes have been predicted to exist at the boundaries of some families of topological superconductors \cite{Qi2009}. TR symmetric topological superconductors have been theoretically proposed to be realized, for example, in Cu$_x$Bi$_2$Se$_3$\citep{Fu_Berg} and in materials with strong spin-orbit coupling in proximity to $\pi$ Josephson junction\cite{Berg2013, Loss2015} or $s_\pm$-wave superconductors\cite{Zhang2013}. Among the expected exotic properties of Majorana Kramers pairs are non-Abelian statistics\cite{Liu2014} and topological two-channel Kondo effect\cite{bao2016}.

The Kramers pair of Majorana modes has also been predicted to appear in TR-invariant vortex core of a TR-invariant topological superconductor\cite{Qi2009}. TR transformation changes the parity of fermion number associated with the zero modes, and therefore acts as a supersymmetry in the vortex\cite{Qi2009}.  However, it remains unclear how a vortex defect that preserves TR symmetry can be physically realized, because a vortex in a superconductor is typically induced by a magnetic field that explicitly breaks TR symmetry. Here we propose that a TR-symmetric vortex naturally appears in a topological superconductor with nematic order.  The nematic superconductor has a two-component order parameter\cite{Fu_Nematic}, which can be viewed as a planar vector. We demonstrate analytically and numerically that a nematic vortex, such that only the vector part of the order parameter rotates, without $U(1)$ phase winding, is TR symmetric and binds a Majorana Kramers pair. 

Our work is motivated by the recent experimental studies of  superconducting $M$-doped Bi$_2$Se$_3$, where the metal element $M$ can be Cu, Nb and Sr.  Shortly after the experimental discovery of superconductivity in Cu$_x$Bi$_2$Se$_3$\cite{Hor2010}, Fu and Berg first proposed that  a TR-invariant topological superconductor could be realized in this system\cite{Fu_Berg}. Majorana surface states associated with the topological superconductor were expected to be observable by point contact measurement\cite{Sasaki2011}; however, scanning tunneling microscopy measurements\cite{Levy2013} performed so far have not detected them. Despite the lack of observed surface states, measurements of bulk properties have revealed compelling evidence that the superconducting state in $M_x$Bi$_2$Se$_3$ spontaneously breaks the lattice discrete rotational symmetry and therefore is nematic, and potentially topological. The main experimental evidence comes from the two-fold anisotropy of various bulk responses to an external magnetic field, including Knight shift\cite{matano2016spin}, specific heat\cite{yonezawa2017}, upper critical field\cite{yonezawa2017,pan2016rotational} and magnetic torque\cite{Asaba2017}.  Corresponding theories of these bulk properties have also been developed\cite{hashimoto2013bulk,Nagai2016,Venderbos2016,kozii2016}.

Our paper is organized as follows. In Sec.~\ref{Sec:continuous_model}, we present the effective Hamiltonian\cite{Venderbos2016_pairing} for the nematic odd-parity superconductor, and review its symmetry properties and gap structure. In Sec.~\ref{continuous_vortex}, we study electronic states in the continuous nematic vortex and provide an analytic solution to the Majorana Kramers pair. The zero modes can be understood in two different limits. In the limit of only helical $p$-wave pairing, spin up (down) electrons form $p_x + i p_y$ ($p_x - i p_y$) Cooper pair condensates. A nematic vortex thus exactly realizes a TR-symmetric vortex discussed in Ref.~\onlinecite{Qi2009}, where spin up and down electrons respectively experience a vortex and antivortex in the pairing order parameters.  In the other limit of only polar $p$-wave pairing, the bulk of the superconductor is gapless. Surprisingly, we find that  a single vortex can induce a flat band of zero energy modes. % whose number scales with the system size.
A finite helical $p$-wave pairing potential lifts all the zero modes to finite energy except for one pair of Majorana modes. In Sec.~\ref{Sec:discrete_vortex}, we study a discrete nematic vortex, which can be experimentally realized at the core where three nematic domains meet.  In agreement with the analysis of continuous vortex, the discrete nematic vortex also binds a Kramers pair of Majorana modes. When the out-of-plane dispersion is included, the Majorana modes evolve into helical modes. Finally in Sec.~\ref{Sec:discussion}, we discuss possible ways to experimentally probe the zero modes in nematic vortex.

\section{Hamiltonian and symmetry}
\label{Sec:continuous_model}
Undoped Bi$_2$Se$_3$ is a strong topological insulator with both TR symmetry and inversion symmetry. Electronic bands in Bi$_2$Se$_3$ are therefore doubly degenerate at each $k$ point. For $M_x$Bi$_2$Se$_3$, we assume that the chemical potential lies in the conduction bands.  When only the conduction bands near $\Gamma$ point are retained in a low-energy theory, the superconducting state in $M_x$Bi$_2$Se$_3$ can be described by a one-orbital model\cite{Fu_Berg, Yip2013,Venderbos2016_pairing}:
\begin{equation}
\begin{aligned}
H& =\frac{1}{2}\int d\boldsymbol{r} \Psi_{\boldsymbol{r}}^\dagger\mathcal{H}(\kk)\Psi_{\boldsymbol{r}},\\
\Psi_{\boldsymbol{r}}^\dagger& = (c_{\boldsymbol{r}\uparrow}^{\dagger}, c_{\boldsymbol{r}\downarrow}^{\dagger}, c_{\boldsymbol{r}\downarrow}, -c_{\boldsymbol{r}\uparrow}),\\
\mathcal{H}(\kk)& = \mathcal{H}_0(\kk)\tau_z + \Delta(\kk)\tau_+ + \Delta^\dagger(\kk) \tau_-. 
\end{aligned}
\label{Hamiltonian}
\end{equation}
Here $\kk$ is the three-dimensional momentum operator, and $\uparrow$ and $\downarrow$ are the pseudo-spin label of the two degenerate states in the conduction band. $\tau_{\pm}=(\tau_x\pm i\tau_y)/2$, and $\tau_{x, y, z}$ are Pauli matrices in Nambu space. $\mathcal{H}_0$ is the kinetic energy arising from band structure. We assume a parabolic dispersion: $\mathcal{H}_0(\kk)=\beta \kk^2 -\mu$, where $\mu>0$ is the chemical potential. The nematic superconductor has an odd-parity pairing in the two-component $E_u$ representation of the $D_{3d}$ point group. The pairing potential can be expressed as:
\begin{equation}
\Delta(\kk) = \boldsymbol{\eta}\cdot \boldsymbol{F}(\kk),
\label{pairing}
\end{equation}
where both $\boldsymbol{\eta}$ and $\boldsymbol{F}$ represent a two-component vector. 
$\boldsymbol{\eta}=(\eta_x, \eta_y)$ acts as the nematic order parameter.
$\boldsymbol{F}$ can be decomposed into three parts when only linear $\kk$ terms are retained\cite{Venderbos2016_pairing}:
\begin{equation}
\begin{aligned}
F_\alpha & = \lambda_1 F_\alpha^{(1)} +\lambda_2 F_{\alpha}^{(2)}+\lambda_3 F_{\alpha}^{(3)}\\
F_{\alpha}^{(1)} & =k_\alpha \sigma_z, \,\,\,\,\,\,\,\,\,\,\,\,\,\,\,\,\,\,\,\,\,
F_{\alpha}^{(3)} =k_z \sigma_\alpha,\\
F_x^{(2)} & = k_x \sigma_y + k_y \sigma_x,\,
F_y^{(2)}  = k_x \sigma_x - k_y \sigma_y,
\end{aligned}
\end{equation}
where $\alpha=x, y$ and $F_{x, y}$ are the two components of $\boldsymbol{F}$. $\sigma_{x, y, z}$ are Pauli matrices in spin space.
$\lambda_{1, 2, 3}$ are real parameters that characterize the relative strength of different pairing terms.
When $\boldsymbol{\eta}$ is real, $\boldsymbol{F}^{(1)}$ and $\boldsymbol{F}^{(3)}$ represent polar $p$-wave pairing, and $\boldsymbol{F}^{(2)}$ is helical $p$-wave pairing where spin up (down) electrons respectively form $p_x + i p_y$ ($p_x - i p_y$) Cooper pairs.

\begin{figure}[t]
	\includegraphics[width=0.95\columnwidth]{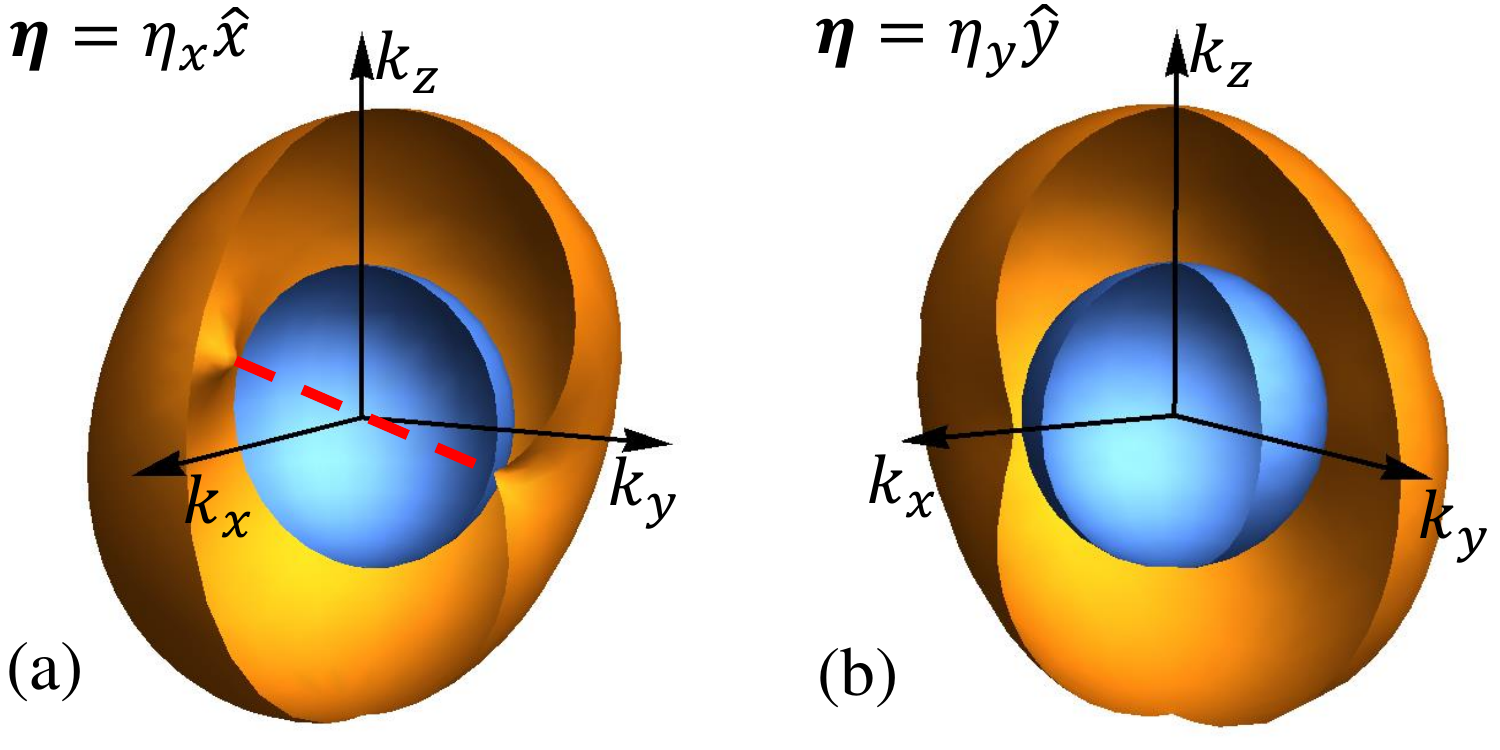}
	\caption{Schematic gap structure (bronze color) on the Fermi surface (blue color). (a) $\boldsymbol{\eta}=\eta_x \hat{x}$. Red dashed line connects the two nodal points. (b) $\boldsymbol{\eta}=\eta_y \hat{x}$. There is a full superconductivity gap.   }
	\label{Fig:gap}
\end{figure}

We now elaborate on the symmetry properties when $\boldsymbol{\eta}$ is spatially uniform.
The pairing potential has odd parity: $\Delta(-\kk)=-\Delta(\kk)$.
TR symmetry is preserved if $\boldsymbol{\eta}$ is real:
$T \mathcal{H}(\kk) T^{-1} = \mathcal{H}(-\kk)$,
where $T = -i \sigma_y K$ is the TR operator and $K$ is the complex conjugate.
%The odd-parity paring would break TR symmetry if $\eta_{1, 2}$ is complex.
The particle-hole (PH) symmetry can be expressed as: $\Xi \mathcal{H}(\kk) \Xi^{-1} =- \mathcal{H}(-\kk)$, where $\Xi=\sigma_y \tau_y K$.

The pairing potential $\Delta$ spontaneously breaks the three-fold rotational symmetry of Bi$_2$Se$_3$.   
Rotating $\boldsymbol{\eta}$ by $2\pi/3$ leads to a different but energetically degenerate pairing potential. 
This is manifested by the following identity:
\begin{equation}
\Delta( \mathcal{R}_3\boldsymbol{\eta}, \mathcal{R}_3\kk_\perp, k_z ) =  \mathcal{C}_3 \Delta( \boldsymbol{\eta}, \kk_\perp, k_z )\mathcal{C}_3^{\dagger},
\end{equation}
where $\kk_\perp=(k_x, k_y)$ is the in-plane momentum. $\mathcal{R}_3$ is the real-space rotation matrix for a $2\pi/3$ rotation around $z$ axis, and $\mathcal{C}_3= \exp(-i \pi \sigma_z/3) $ is the rotation matrix in spin space.

In  the $D_{3d}$ point group there are three mirror planes that are located at $x=0$ or its three-fold rotation counterparts. A mirror symmetry ($x \leftrightarrow -x $) is preserved if $\boldsymbol{\eta}$ is perpendicular to the corresponding mirror plane $x=0$:
\begin{equation}
\begin{aligned}
&\Delta(\boldsymbol{\eta}=\eta_x\hat{x}, -k_x, k_y, k_z)\\
= &\mathcal{M} \Delta(\boldsymbol{\eta}=\eta_x\hat{x}, k_x, k_y, k_z)  \mathcal{M}^\dagger,
\end{aligned}
\label{mirror}
\end{equation}
where $\mathcal{M}=\exp(-i \pi \sigma_x/2)$ is the mirror operator in spin space.
The superconductor has point nodes protected by the mirror symmetry when $\boldsymbol{\eta}$ is along $\hat{x}$\cite{Zhang_Mirror, Zhang_2014, Fu_Nematic}, as we discuss below.

As an odd-parity pairing potential, $\Delta$ can be expressed in terms of a vector $\boldsymbol{d}$:
\begin{equation}
\Delta(\kk)=\boldsymbol{d}(\kk)\cdot \boldsymbol{\sigma}.
\end{equation}
When $\boldsymbol{\eta}$ is real, the superconductivity gap is given by $|\boldsymbol{d}(\kk)|$ on the Fermi surface\cite{Sigrist_Ueda}. 
We present the explicit form of $\boldsymbol{d}(\kk)$ for two representative cases. 
In the case of  $\boldsymbol{\eta}=\eta_x \hat{x}$, 
\begin{equation}
\boldsymbol{d}(\kk) = \eta_x (\lambda_2 k_y + \lambda_3 k_z, \lambda_2 k_x, \lambda_1 k_x). 
\end{equation}
Because of the mirror symmetry in (\ref{mirror}), $d_{y, z}$ vanishes on the mirror plane $k_x=0$. 
Since $\boldsymbol{d}$ is odd under inversion, $|\boldsymbol{d}(\kk)|$ must have nodes on the mirror plane. 
The point nodes are located at:
\begin{equation}
\kk=\pm(0, \lambda_3, -\lambda_2) k_F/\sqrt{\lambda_3^2+\lambda_2^2},
\end{equation}
where $k_F$ is the Fermi wave vector $\sqrt{\mu/\beta}$.
Note that the point nodes appear at non-zero $k_z$.

In another case $\boldsymbol{\eta}=\eta_y \hat{y}$,
\begin{equation}
\boldsymbol{d}(\kk) = \eta_y (\lambda_2 k_x, -\lambda_2 k_y+\lambda_3 k_z, \lambda_1 k_y).
\end{equation}
It is then impossible to make all three components of $\boldsymbol{d}(\kk)$ to vanish simultaneously on the Fermi surface, so the superconductor is fully gapped. The gap structures for the two different cases are schematically shown in Fig.~\ref{Fig:gap}. In summary, Hamiltonian (\ref{Hamiltonian}) describes a topological Dirac superconductor\cite{Zhang_2014} with surface Majorana arcs when $\boldsymbol{\eta}$ is perpendicular to one of the mirror planes, and otherwise a fully gapped odd-parity topological superconductor\citep{Fu_Berg}.

\section{Continuous nematic vortex}
\label{continuous_vortex}

\begin{figure}[t]
	\includegraphics[width=1\columnwidth]{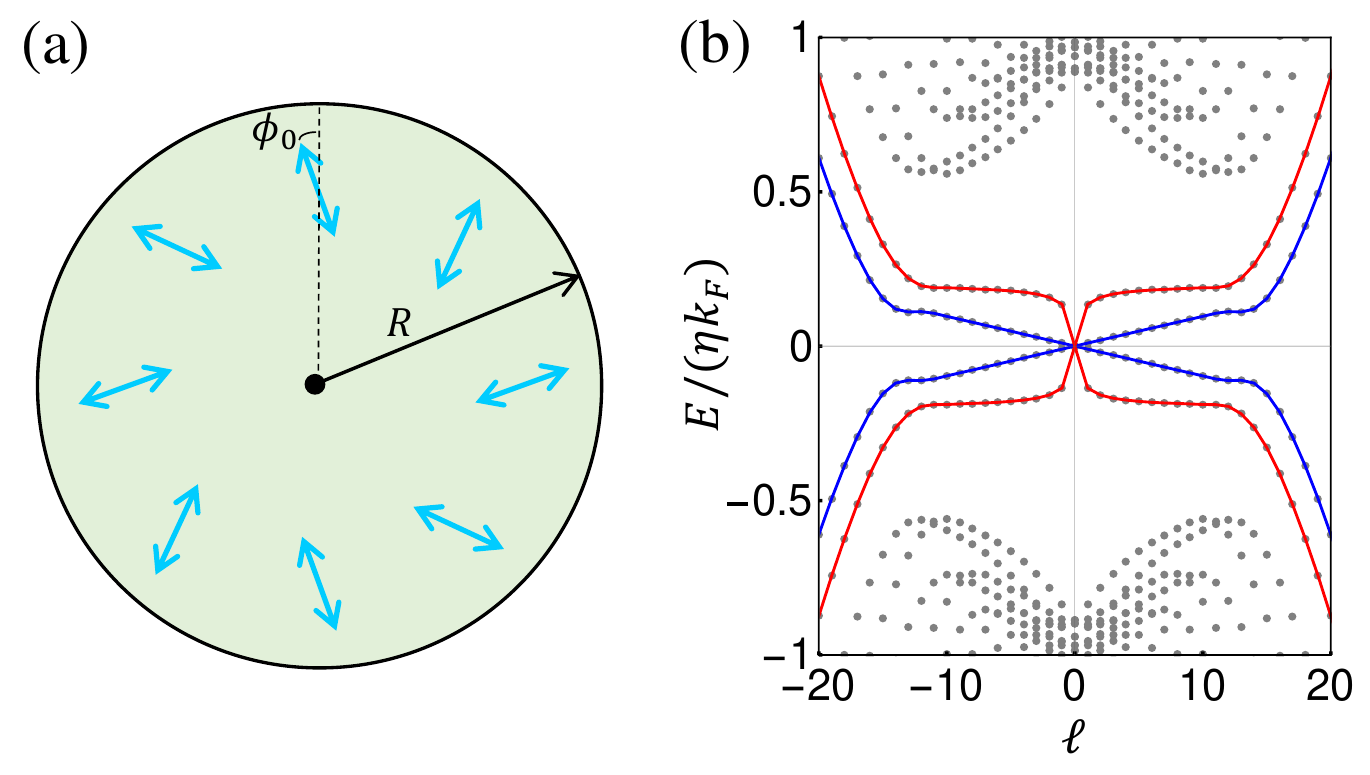}
	\caption{ (a) Illustration of a nematic vortex. The arrows indicate the orientation of the nematic order parameter $\boldsymbol{\eta}$. $\phi_0$ describes the angle between  $\boldsymbol{\eta}$ and the radial direction. We set $\phi_0$ to be zero unless otherwise stated. (b) The energy spectrum of a {\it gapped} superconductor with a nematic vortex. The Hamiltonian is specified in (\ref{polar}) The calculation was performed for a disk with a radius $k_F R=20$. To reduce computational cost, we used a large pairing potential $\eta k_F/\mu=1$ for all numerical calculations in the paper. $\lambda_1$ and $\lambda_2$ are respectively taken to be 1 and 0.2. Red and blue lines respectively mark states localized near the core center and the outer edge of the disk. There are four modes that have nearly zero energy ($\sim 2\times 10^{-5}\mu$) at $\ell=0$. Proper linear combination of these four modes results in two states localized near the core, and another two states near the edge.  }
	\label{Fig:gappedvortex}
\end{figure}

\begin{figure*}[t]
	\includegraphics[width=2\columnwidth]{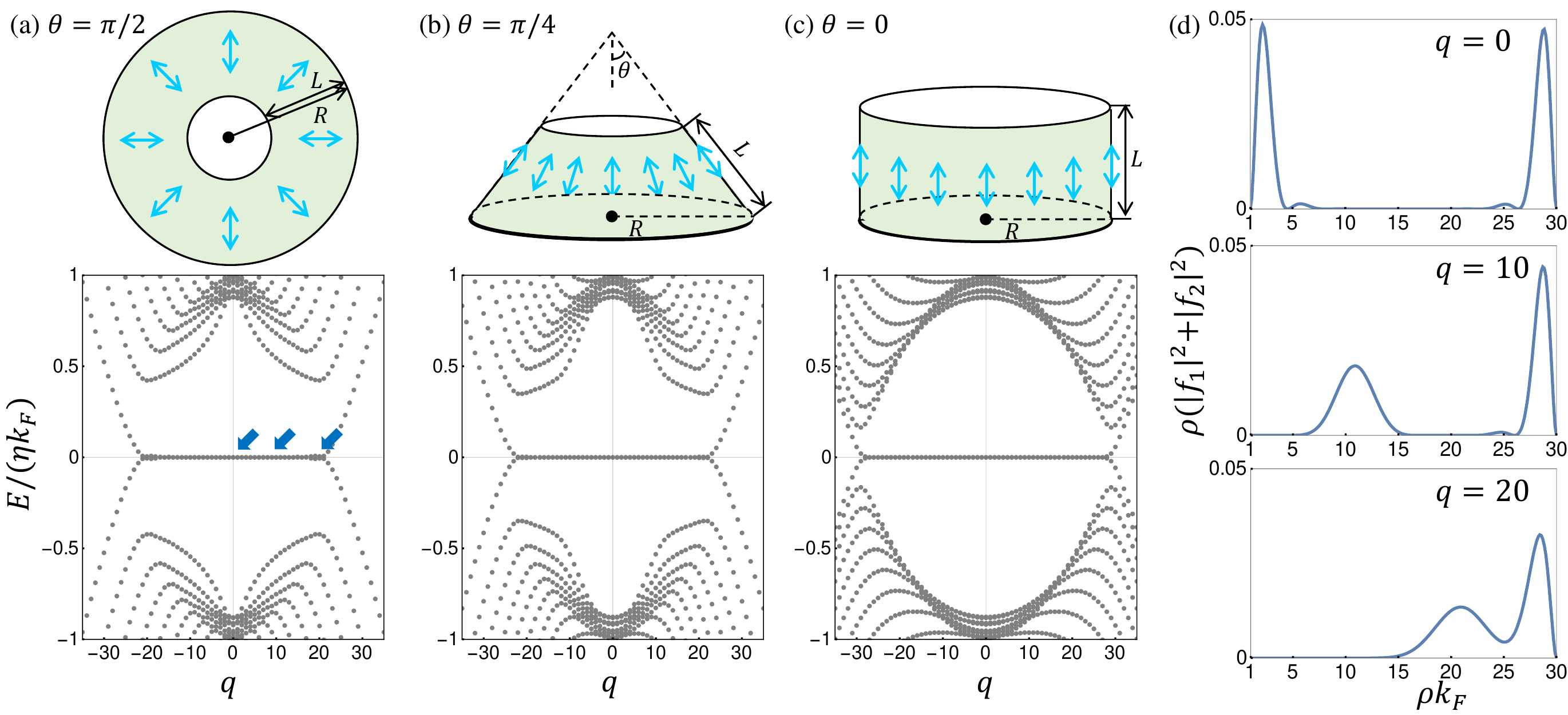}
	\caption{(a)-(c) The lower panels show the energy spectrum of a polar $p$-wave superconductor ($\lambda_1=1$, $\lambda_2=0$ ) described by (\ref{nonchiral_p}) in different geometries illustrated in the corresponding upper panel. (a) The nematic order parameter in the plane geometry ($\theta=\pi/2$) has a vortex structure. (b) When $\theta$ is less than $\pi/2$, Hamiltonian (\ref{nonchiral_p}) can be viewed as defined on a truncated cone, where the radial coordinate $\rho$ is measured from the tip of the cone. (c) When $\theta$ approaches $0$, the truncated cone becomes a cylinder and the nematic order parameter becomes spatially uniform. Parameters $R$ and $L$ are kept fixed when $\theta$ varies: $k_F R=30$ and $k_F L=29$. (d) Real space probability distribution of states  that have an energy slightly above zero at $q=$0, 10 and 20 in the plane geometry ($\theta=\pi/2$). The states are indicated by the blue arrows in (a).	 }
	\label{Fig:cone}
\end{figure*}

Since the nematic superconductor has a two-component order parameter, 
a continuous vortex defect along $\hat{z}$ is characterized by two winding numbers $(n, m)$:
\begin{equation}
\boldsymbol{\eta}(\boldsymbol{r}) = \eta(\rho) e^{i n \phi} ( \cos m \phi ,\, \sin m \phi ),
\label{vortex}
\end{equation}
where $(\rho, \phi)$ are the polar coordinates of the in-plane position $\boldsymbol{r}_\perp$.
The usual superconducting vortex has an integer phase winding number $n$ and a zero value of $m$.
$n$ and $m$ can also take half integer values: $(n, m)=(\pm 1/2, \pm 1/2)$,
which correspond to half vortices\cite{Ivanov2001}. Vortices with finite $n$  break TR symmetry. 
In this paper, we focus on a TR-invariant vortex with $(n, m)= (0, 1)$, which we dub {\em nematic vortex}. 
%The parameter $\phi_0$ in (\ref{vortex}) is the angle between the order parameter $\boldsymbol{\eta}$ and the radial direction along $\boldsymbol{r}_\perp%$, as illustrated in Fig.~\ref{Fig:gappedvortex}(a).

We study electronic states induced by the nematic vortex. As $k_z$ remains a good quantum number, we start by considering $k_z=0$. The dispersion of vortex states as a function of $k_z$ will be discussed in Sec.~\ref{Sec:discrete_vortex}. The Hamiltonian (\ref{Hamiltonian}) at $k_z=0$ and in the presence of nematic vortex can be expressed in polar coordinates as:
\begin{equation}
\begin{aligned}
\tilde{\mathcal{H}}&=h_0\tau_z+ \big[\lambda_1 h_1 \sigma_z  + \lambda_2 (h_2\sigma_+ + h_2^\dagger \sigma_-)\big]\tau_x,\\
h_0&\equiv \beta \kk_\perp^2-\mu = -\beta(\partial_\rho^2+\frac{1}{\rho}\partial_\rho+\frac{1}{\rho^2 }\partial_\phi^2)-\mu,\\
h_1&\equiv (\boldsymbol{\eta}\cdot \kk + \kk \cdot \boldsymbol{\eta} )/2
=-i \eta (\partial_\rho+\frac{1}{2\rho}) ,\\
h_2&\equiv -i(\eta_+ k_+ + k_+ \eta_+)/2\\
&= -i\eta e^{i 2 \phi}[-i(\partial_\rho-\frac{1}{2\rho})+\frac{\partial_\phi}{\rho}],
\end{aligned}
\label{polar}
\end{equation}
where $\eta=|\boldsymbol{\eta}(\boldsymbol{r})|$, $k_\pm=k_x\pm ik_y$ and $\eta_\pm=\eta_x \pm i\eta_y$.
In (\ref{polar}), the product between $\boldsymbol{\eta}$ and $\kk$ is symmetrized as they no longer commute when $\boldsymbol{\eta}$ has spatial variation. In the nematic vortex, $\eta_+$ and $\eta_-$ respectively has  a vortex $\exp(i\phi)$ and antivortex $\exp(-i\phi)$ structure. 
For the TR invariant vortex that we study, $\eta_+$ and $\eta_-$ vanish at the same location in the vortex core.
In the following we will neglect the spatial variation in  $|\boldsymbol{\eta}(\boldsymbol{r})|$ to simplify calculation.

$\tilde{\mathcal{H}}$ preserves both TR and PH symmetries. %: $T\tilde{\mathcal{H}}T^{-1}=\tilde{\mathcal{H}}$ and $\Xi\tilde{\mathcal{H}}\Xi^{-1}=-\tilde{\mathcal{H}}$. 
Therefore, the nematic vortex described by (\ref{polar}) is in class DIII and can be characterized by a $Z_2$ invariant that signals the presence or absence of Majorana Kramers doublet in the vortex core\cite{Teo2010}. Here the existence of the zero modes is hinted by the limit in which $h_1$ term is absent in (\ref{polar}).  In this limit, Hamiltonian (\ref{polar}) effectively describes a TR symmetric superconductor in which spin up and down electrons respectively form $p_x+ip_y$ and $p_x-ip_y$ Cooper pairs, and the superconducting phase has a vortex (antivortex) defect in the spin up (down) sector. 
Therefore, there must be a Kramers pair of Majorana modes when $\lambda_1=0$.
By explicitly solving the eigenenergy problem of (\ref{polar}), we show that the zero modes are robust regardless of the exact values of $\lambda_{1, 2}$.

Hamiltonian (\ref{polar}) has eigenvectors of the form:
\begin{equation}
e^{i\ell \phi}[e^{i\phi}f_1(\rho),e^{-i\phi}f_2(\rho),e^{i\phi}f_3(\rho),e^{-i\phi}f_4(\rho)]^{\text{T}},
\end{equation}
where $\ell$ takes integer values. 
For numerical diagonalization, $f_i(\rho)$ is expanded in terms of orthonormal Bessel functions on a finite-radius disk\cite{Numeric_Vortex, Hu2015}.
The energy spectrum as a function of $\ell$ is illustrated in Fig. \ref{Fig:gappedvortex}(b).
The numerical results indicate that a pair of zero-energy modes with $\ell=0$ are localized in the vortex core.
Because of PH and TR symmetries, they must be Majorana modes and TR partners:
\begin{equation}
\begin{aligned}
\gamma_{\uparrow}^\dagger&=\int d\boldsymbol{r} [(e^{i\phi}\xi_1(\rho)c_{\boldsymbol{r}\uparrow}^\dagger+e^{-i\phi}\xi_2(\rho)c_{\boldsymbol{r}\downarrow}^\dagger)+h.c.], \\
\gamma_{\downarrow}^\dagger&=\int d\boldsymbol{r} [(-e^{i\phi}\xi_2^*(\rho)c_{\boldsymbol{r}\uparrow}^\dagger+e^{-i\phi}\xi_1^*(\rho)c_{\boldsymbol{r}\downarrow}^\dagger)+h.c.],
\end{aligned}
\label{MKP}
\end{equation}
which satisfy the Majorana condition ($\gamma_{\uparrow, \downarrow}^\dagger=\gamma_{\uparrow, \downarrow}$) and form Kramers doublet under TR transformation ($\gamma_{\uparrow} \rightarrow \gamma_{\downarrow}$, $\gamma_{\downarrow} \rightarrow  - \gamma_{\uparrow}$).
As shown in Appendix \ref{appA},wave functions $\xi_{1,2}$ can be obtained analytically:
\begin{equation}
\begin{aligned}
\big[\xi_1, \xi_2\big]&=\big[i(\lambda_2+\sqrt{\lambda_1^2+\lambda_2^2}), \lambda_1 \big]\xi,\\
\xi(\rho)&=\mathcal{N} J_1(\sqrt{1-(\tilde{\eta}k_F/2\mu)^2}k_F\rho)\exp(-\frac{\tilde{\eta}}{2\beta}\rho),
\end{aligned}
\label{analytic}
\end{equation}
where $\mathcal{N}$ represents a normalization factor and $J_1$ is Bessel function of the first order.
In (\ref{analytic}), $\lambda_{1, 2} \geq 0$ is assumed,
and $\tilde{\eta}$ is the effective total pairing strength $\eta\sqrt{\lambda_1^2+\lambda_2^2}$.

We analyze (\ref{analytic}) in two different limits. In the limit $\lambda_1=0$, $\xi_2$ vanishes and $\gamma_{\uparrow, \downarrow}^\dagger$ is the Majorana mode in corresponding spin sector, as expected. In the  other limit $\lambda_2=0$, the zero modes are also exponentially confined to the vortex center.
This result is surprising, since the Hamiltonian (\ref{Hamiltonian}) with zero $\lambda_2$  describes a {\it gapless} polar $p$-wave superconductor. 
Next we reveal another unexpected effect -- that the polar $p$-wave superconductor supports not just one pair, but a flat band of zero modes localized around the nematic vortex core, despite the bulk being gapless.

Hamiltonian (\ref{Hamiltonian}) with $k_z=0$ and $\lambda_2=0$ reduces to
\begin{equation}
H_1  = \int d\boldsymbol{r} (c_{\boldsymbol{r}\uparrow}^{\dagger}, c_{\boldsymbol{r}\downarrow})\mathcal{H}_1\begin{pmatrix}c_{\boldsymbol{r}\uparrow}\\c_{\boldsymbol{r}\downarrow}^\dagger\end{pmatrix},
\mathcal{H}_1=h_0 \tau_z + \lambda_1 h_1 \tau_x,
\label{nonchiral_p}
\end{equation}
where $h_{0, 1}$ is provided in (\ref{polar}). We reiterate their expressions below but introduce a new parameter $\theta$ that can be used to tune the system geometry,
\begin{equation}
\begin{aligned}
h_0&\equiv \beta \kk_\perp^2-\mu	 = -\beta(\partial_\rho^2+\frac{1}{\rho}\partial_\rho+\frac{1}{\rho^2 \sin^2\theta}\partial_\phi^2)-\mu,\\
h_1&\equiv (\boldsymbol{\eta}\cdot \kk+\kk\cdot \boldsymbol{\eta} )/2 = -i \eta (\partial_\rho+\frac{1}{2\rho}).
\end{aligned}
\label{cone}
\end{equation} 
$\theta=\pi/2$ corresponds to a flat plane geometry. When $\theta<\pi/2$, Eq.~(\ref{cone}) generalizes the problem on a plane to a cone with an opening angle $2\theta$, as illustrated in Fig.~\ref{Fig:cone}. This generalization is derived in Appendix \ref{appB} and is used to shed light on relationship between the vortex core states and edge modes in cylindrical geometry.

Similar to $\tilde{\mathcal{H}}$ in (\ref{polar}), the operator $\mathcal{H}_1$ in (\ref{cone}) has eigenstates of the form
\begin{equation}
\exp(i q \phi)[f_1(\rho), f_2(\rho)]^{\text{T}}, 
\label{ansatz}
\end{equation}
where $q$ is an integer. We solve  (\ref{nonchiral_p}) and (\ref{cone}) in the following range of $\rho$,
\begin{equation}
\frac{R}{\sin\theta}-L<\rho<\frac{R}{\sin\theta},
\label{boundary}
\end{equation}
with open boundary conditions. This specifies an annulus in the plane when $\theta=\pi/2$ and a truncated cone with a base radius $R$ when $\theta<\pi/2$. 
%Both $R$ and $L$ are kept fixed when $\theta$ varies.

An important special case is $\theta\to0$. In this limit, $\rho\to\infty$, while $\rho\sin\theta$ stays at a finite value $R$. Geometrically, the truncated cone becomes a cylinder with a radius $R$ and a finite length $L$. % and $(\rho, R\phi)$ becomes the orthogonal coordinates on the cylinder.
Eq.~(\ref{cone}) then describes  exactly a {\it uniform} polar $p$-wave superconductor on the cylinder, as depicted in Fig.~\ref{Fig:cone}(c).
It is known that a polar $p$-wave superconductor supports a flat band of zero-energy modes confined to edges\cite{Martin2012}.

Equations (\ref{cone}) and (\ref{boundary}) state that the polar $p$-wave superconductor with a nematic vortex in the annular geometry can be related to a {\it uniform} one on the cylinder via a single parameter $\theta$.  We solved the eigenvalue problem for different $\theta$ values using the  ansatz  (\ref{ansatz}). The energy spectrum as a function of  $q$ is shown in Fig. \ref{Fig:cone}.  The zero-energy modes with small $|q|$ index on cylinder ($\theta=0$) remain at zero as $\theta$ increases, all the way to $\theta=\pi/2$. 

Because of a PH symmetry of $\mathcal{H}_1$ ($\tau_y \mathcal{H}_1=-\mathcal{H}_1 \tau_y$), the zero modes are eigenstates of $\tau_y$.
Therefore the $q$th zero mode localized near the inner edge can be parametrized as:
\begin{equation}
d_q^\dagger=\int d\boldsymbol{r} e^{iq\phi} \psi_q(\rho)[i c_{\boldsymbol{r}\uparrow}^\dagger+c_{\boldsymbol{r}\downarrow}],
\label{dq}
\end{equation}
where $\psi_q(\rho)$ is the radial part of the wave function.
The degree of confinement of $\psi_q(\rho)$ to the inner edge depends on the index $q$, as illustrated in Fig.~\ref{Fig:cone}(d).
%When $\phi_0=0$, the nematic order parameter $\boldsymbol{\eta}$ is parallel to the radial direction, providing a confinement potential.
%As $\phi_0$ increases, the confinement along radial direction becomes weaker. 
%In the special case of $\phi_0=\pi/2$, the confinement is completely lost and there are no zero-energy modes, as mentioned above for the cylinder case.
$\psi_q(\rho)$ moves away from the inner edge as $|q|$ increases due to a ``centrifugal force" in (\ref{cone}). 
In the annulus geometry ($\theta=\pi/2$), the zero modes are robust when the inner radius becomes zero ($L\rightarrow R$). As shown in Appendix \ref{appC},  an analytic solution of the wave function $\psi_q(\rho)$ can be obtained at every integer $q$ for a disk with an infinite radius. The number of zero modes becomes finite on a finite-radius disk due to hybridization between modes deriving from the core and the outer edge.
The stability of the flat band with respect to disorder and interactions is an interesting topic, beyond the scope of this paper. Previous work on related problems can be found, for example, in Ref \onlinecite{Li2013,Potter2014, Baum2015}. 

%An important property of the zero modes is revealed by TR transformation $T$:
%\begin{equation}
%T d_q^\dagger T^{-1}= -i d_q,\,\, T d_q T^{-1} = i d_q^\dagger.
%\end{equation}
%Here TR transformation changes fermion parity and therefore acts as a supersymmetry.
%%The fermion number operator $d_q^\dagger d_q$ breaks TR symmetry.
%Note that $d_q^\dagger$ represents a fermion mode instead of a Majorana mode because $d_q^\dagger\neq d_q$.

Starting from the polar phase, we can examine the effect of the helical part of the pairing potential $\lambda_2 (\boldsymbol{\eta}\cdot\boldsymbol{F}^{(2)}+\boldsymbol{F}^{(2)}\cdot \boldsymbol{\eta})/2$ by including it as a perturbation, and projecting it onto the zero modes $d_q^\dagger$. The projection leads to the following effective low energy Hamiltonian:
\begin{equation}
\tilde{H}_{2}=\lambda_2\sum_{p>0}\varepsilon_p d_{1+p}^\dagger d_{1-p}^\dagger+\varepsilon_p^* d_{1-p} d_{1+p},
\end{equation}
where $p=0$ term is absent because of fermion anticommutation relation. 
%Also note that $d_1^\dagger d_1$ breaks time reversal symmetry and is absent.
Therefore, all zero modes except $d_1^\dagger$ are lifted to finite energy when $\lambda_2\neq 0$.
$d_1^\dagger$ can be decomposed into two Majorana modes:
\begin{equation}
\gamma_\uparrow^\dagger=d_1^\dagger+d_1, \gamma_\downarrow^\dagger= i(d_1^\dagger-d_1),
\end{equation}
which are exactly those presented in (\ref{MKP}).
As a summary, the Majorana Kramers doublet $\gamma_{\uparrow,\downarrow}^\dagger$ can be understood either from the helical limit $\lambda_1=0$ or the polar limit $\lambda_2=0$.

The degeneracy between $\gamma_{\uparrow}^\dagger$ and $\gamma_{\downarrow}^\dagger$ is protected by TR symmetry.
TR changes the fermion parity associated with fermion mode $d_1$ and acts as: $d_1^\dagger \rightarrow -i d_1$ and $d_1 \rightarrow i d_1^\dagger$,
which is a manifestation of supersymmetry.  A mass term $d_1^\dagger d_1$ is prohibited by TR symmetry.
When TR symmetry is explicitly broken, for example by a Zeeman term of the form $\epsilon_{\text Z} \sigma_z$, generally there can be coupling between $\gamma_{\uparrow}^\dagger$ and $\gamma_{\downarrow}^\dagger$ that lifts the degeneracy.

\begin{figure*}[t!]
	\includegraphics[width=1.9\columnwidth]{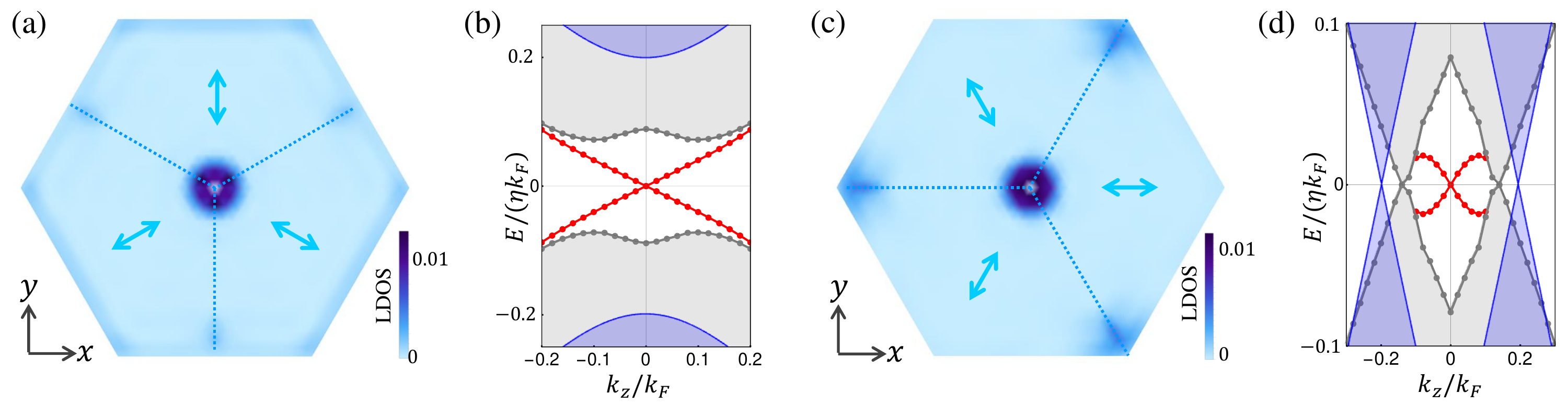}
	\caption{(a) and (b) Type I discrete nematic vortex. (a) Real space local density of states (LDOS) corresponding to one of the four modes  at $k_z=0$ that have nearly zero energy ($\sim 5\times 10^{-4}\mu$). The three dashed lines mark the boundaries between different nematic domains, which also coincide with the three mirror planes.  The nematic order parameter is uniform within each domain and its direction is illustrated by the arrow. (b) The red lines indicate the $k_z$ dispersion of states localized near the core center. The shaded region illustrates the bulk excitation continuum. The blue lines show the minimum energy excitation of a {\it uniform} state with $\boldsymbol{\eta}$ along $y$ direction. Parameter values are $(\lambda_1, \lambda_3, \lambda_2)=(1, 1, 0.2)$. (c) and (d) Corresponding plots for type II discrete nematic vortex. In (d) the blue lines show the minimum energy excitation of a {\it uniform} state with $\boldsymbol{\eta}$ along $x$ direction.}
	\label{Fig:lattice}
\end{figure*} 

\section{Discrete nematic vortex}
\label{Sec:discrete_vortex}

Because Bi$_2$Se$_3$ only has a discrete three-fold rotational symmetry, the nematic vector $\boldsymbol{\eta}$ is expected to be pinned to one of the high symmetry directions. The Ginzburg-Landau free energy for a uniform state is\cite{Fu_Nematic}:
\begin{equation}
\begin{aligned}
\mathcal{F} = &\,\,\,\,\,\,\, b_0(|\eta_x|^2+|\eta_y|^2)+b_1 (|\eta_x|^2+|\eta_y|^2)^2\\
&+b_2(\eta_x^2+\eta_y^2)^2
+ b_3 [(\eta_+^*\eta_-)^3+(\eta_-^*\eta_+)^3].
\end{aligned}
\end{equation}
Here the sign of $b_2$ distinguishes between nematic and chiral states, which have respectively real and complex order parameters $\boldsymbol{\eta}$.
We assume $b_2<0$ so that a nematic state is energetically favorable. 
The sixth-order terms in $\mathcal{F}$ capture the crystalline anisotropy. 
The sign of $b_3$ determines whether $\boldsymbol{\eta}$ is pinned to be perpendicular or parallel to one of the three mirror planes, for example, along $x$ or $y$ direction.
Because of the lattice anisotropy, a continuous nematic vortex described by (\ref{vortex}) is energetically unfavorable. Nevertheless, a discrete vortex can be realized around the core where three nematic domains meet, as illustrated in Fig.~\ref{Fig:lattice}. 
Indeed, a recent experiment on the specific heat of Cu$_x$Bi$_2$Se$_3$ under magnetic field has indicated the existence of multiple nematic domain walls within one sample\cite{yonezawa2017}. Therefore, a discrete nematic vortex may naturally appear due to local disorder or strain effects.

To numerically study the discrete vortex, we regularize the Hamiltonian (\ref{Hamiltonian}) on a triangular lattice. Two representative types of discrete vortex are studied. In type I, $\boldsymbol{\eta}$ within each domain is parallel to a mirror plane. The bulk of each domain is therefore fully gapped, as illustrated in Fig.~\ref{Fig:gap}(b).   
The numerical calculation for $k_z=0$ indicates that there are four zero modes. Two of them are localized around the core center and the other two are bound to edges[see Fig.~\ref{Fig:lattice}(a)], which are expected from the analysis of the continuous vortex. At finite $k_z$, the two zero modes localized around the core evolve into two branches that disperse linearly with $k_z$ as shown in Fig.~\ref{Fig:lattice}(b) and therefore can be identified as helical Majorana modes. 

In type II,  $\boldsymbol{\eta}$ within each domain is perpendicular to a mirror plane. The bulk of each domain in this case is gapless, and the point nodes are at $k_z\neq 0$,  as shown in Fig.~\ref{Fig:gap}(a). There are still two zero modes localized around the core for $k_z=0$, at which the bulk is fully gapped.  At finite $k_z$, the helical modes merge into the bulk excitation continuum.
The bulk excitations of the discrete vortex structure become gapless at a finite $k_z$ (Fig.~\ref{Fig:lattice}(d)). The exact values of $k_z$ where the bulk excitation minimum are located are slightly different for the domain structure and the uniform state due to the presence of domain core and finite size effects.
As $k_z$ may not be conserved in the presence of disorder, the zero modes in type II vortex are not as robust as those in type I.

The energy cost of a domain wall scales linearly with the area of the interface between two domains. This energy cost could be remedied by studying a sample with mesoscopic geometries\cite{Kim2007}. Another possibility that discrete vortices could be stabilized is due to Imry-Ma scenario\cite{Imry_Ma}, as the nematic order parameter can couple to local disorder or strain field\cite{Fu_Nematic}.

\section{Discussion}
\label{Sec:discussion}
The presence of the Majorana Kramers pair will lead to an enhanced density of states (DOS) near the vortex core at zero energy. To probe the enhanced DOS, it would be favorable to study materials in a thin film geometry so that the motion along $z$ is quantized. Scanning tunneling microscope (STM) is particularly suitable to search for the zero modes. STM has been successfully used to detect Majorana modes in different schemes\cite{nadj2014,Sun2016}, and to identify nematic states in the quantum Hall regime\cite{Feldman2016}. 
Within each nematic domain of $M_x$Bi$_2$Se$_3$, the nematic order parameter will lead to a two-fold anisotropy in electronic band structure, which should be detectable in quasiparticle interference pattern in STM\cite{Hofmann2013, Farrell2015, Lambert2017}. Different nematic domains will have different interference pattern, and thus can also be distinguished.   

In summary, a nematic vortex  can be realized in odd-parity topological superconductors. It is an unusual type of topological defect that  preserves time reversal symmetry and binds a Kramers pair of Majorana modes. Other interesting properties of nematic odd-parity topological superconductors are likely to await discovery.

\section{Acknowledgment}
We acknowledge support
from Department of Energy, Office of Basic Energy
Science, Materials Science and Engineering Division.

\appendix
\section{Analytic solution to zero modes in nematic vortex}
\label{appA}
In this appendix, we present the details of obtaining analytic solution to Majorana zero modes.
Equation (\ref{MKP}) is an ansatz to zero modes of Hamiltonian (\ref{polar}). Wave functions $\xi_{1,2}$ satisfy the following equations:
\begin{equation}
\begin{aligned}
\tilde{h}_0\xi_1+\lambda_1 h_1\xi_2^*-\lambda_2\tilde{h}_2\xi_1^*=0,\\
\tilde{h}_0\xi_2^*-\lambda_1 h_1\xi_1-\lambda_2\tilde{h}_2\xi_2=0,
\end{aligned}
\label{UHU}
\end{equation}
where $\tilde{h}_0$ and $\tilde{h}_2$ are given by:
\begin{equation}
\begin{aligned}
\tilde{h}_0=&e^{-i\phi}h_0e^{i\phi}=-\beta(\partial_\rho^2+\frac{1}{\rho}\partial_\rho-\frac{1}{\rho^2 })-\mu,\\
\tilde{h}_2=&e^{-i\phi}h_2e^{-i\phi}=-\eta(\partial_\rho+\frac{1}{2\rho}).
\end{aligned}
\end{equation}
It is important to note that $h_1=i \tilde{h}_2$, which makes analytic solution possible.  Equation (\ref{UHU}) reduces to:
\begin{equation}
\begin{aligned}
\tilde{h}_0\xi_1+\tilde{h}_2(i \lambda_1 \xi_2^*-\lambda_2\xi_1^*)=0,\\
\tilde{h}_0\xi_2^*-\tilde{h}_2(i\lambda_1 \xi_1+\lambda_2\xi_2)=0.
\end{aligned}
\label{UHU_1}
\end{equation}
To proceed we make the following ansatz:
\begin{equation}
\begin{aligned}
&\big(\xi_1, \xi_2\big)=\big(u, v \big)\xi,\\
&\xi=\mathcal{N} J_1(\sqrt{1-(\Lambda \eta k_F/2\mu)^2}k_F\rho)\exp(-\frac{\Lambda \eta}{2\beta}\rho),
\end{aligned}
\label{ansatz_xi}
\end{equation}
where $u$, $v$ and $\Lambda$ are parameters that need to be determined. $\mathcal{N}$ is just a normalization factor. 
Ansatz in (\ref{ansatz_xi}) is proposed based on the following identity:
\begin{equation}
(\tilde{h}_0+\Lambda \tilde{h}_2)\xi=0.
\end{equation}
Equation (\ref{UHU_1}) then simplifies to an algebraic equation:
\begin{equation}
\begin{aligned}
-\Lambda u+ i\lambda_1 v^*-\lambda_2 u^*=0,\\
\Lambda v^* + i\lambda_1 u+\lambda_2 v=0.
\end{aligned}
\label{uvL}
\end{equation}
When $u$ and $v$ are decomposed into their real and imaginary parts, equation (\ref{uvL}) can be viewed as an eigenproblem with $\Lambda$ as the eigenvalue. There are two eigenvalues $\pm\sqrt{\lambda_1^2+\lambda_2^2}$. We choose $\Lambda=\sqrt{\lambda_1^2+\lambda_2^2}$ so that the zero modes are exponentially confined to the center. Given $\Lambda=\sqrt{\lambda_1^2+\lambda_2^2}$, there are two solutions for $(u, v)$:
\begin{equation}
\begin{aligned}
(u_1, v_1)&=\big[i(\lambda_2+\sqrt{\lambda_1^2+\lambda_2^2}), \lambda_1 \big],\\
(u_2, v_2)&=\big(-v_1,u_1^*\big),
\end{aligned}
\end{equation}
where $\lambda_{1, 2}\geq 0$ is assumed without loss of generality.
$(u_1, v_1)$ leads to the solution in (\ref{analytic}). $(u_2, v_2)$ does not give rise to an independent solution but just switches the definition of $\gamma_\uparrow^\dagger$ and $\gamma_\downarrow^\dagger$.

\section{Cone geometry}
\label{appB}
In this appendix, we present some mathematical properties on the cone geometry. The coordinate on a cone is $(\rho, \phi)$, where $\rho$ is the radial coordinate measured from the tip of the cone, and $\phi$ is the azimuthal angle. The gradient and Laplace operators on the cone are
\begin{equation}
\begin{aligned}
\nabla&=\partial_\rho \hat{\rho}+\frac{1}{\rho \sin\theta}\partial_\phi \hat{\phi},\\
\nabla^2&=\partial_\rho^2+\frac{1}{\rho}\partial_\rho+\frac{1}{\rho^2 \sin^2\theta}\partial^2_\phi,
\end{aligned}
\end{equation}
where $\hat{\rho}$ and $\hat{\phi}$ are respectively the unit vectors along the radial and the azimuthal direction.
$\theta$ is half of the opening angle of the cone, as shown in Fig.~\ref{Fig:cone}(b).
The following identities are used to derive the Laplace operator from the gradient operator:
\begin{equation}
\partial_\rho \hat{\phi}=0, \,\, \partial_\phi \hat{\rho}= \sin\theta \hat{\phi}.
\end{equation}

We now show how the Hamiltonian in ({\ref{nonchiral_p}}) and (\ref{cone}) are derived on the cone geometry.
The kinetic energy has a simple connection to the Laplace operator: $h_0=\beta k_\perp^2-\mu=-\beta \nabla^2-\mu$.
The pairing order parameter  $\boldsymbol{\eta}$ varies in space in the following way:
\begin{equation}
\boldsymbol{\eta}=\eta(\cos\phi_0 \hat{\rho}+ \sin\phi_0 \hat{\phi}),
\end{equation}
where $\phi_0$ is the angle between the order parameter $\boldsymbol{\eta}$ and the radial direction.
In the main text, we have set $\phi_0=0$.  
$h_1$ has the expression:
\begin{equation}
\begin{aligned}
h_1&=-i(\boldsymbol{\eta}\cdot\nabla+\nabla\cdot\boldsymbol{\eta})/2\\
&=-i ( \eta \cos\phi_0 )(\partial_\rho+\frac{1}{2\rho}) -i\frac{\eta \sin\phi_0}{\rho \sin\theta}\partial_\phi.
\end{aligned}
\end{equation}

When $\phi_0=0$, the nematic order parameter $\boldsymbol{\eta}$ is parallel to the radial direction, providing a confinement potential.
As $\phi_0$ increases, the confinement along radial direction becomes weaker. 
In the special case of $\phi_0=\pi/2$, the confinement is completely lost and there are no zero-energy modes.
Therefore, the presence of zero modes in the nematic vortex of a polar $p$-wave superconductor requires that $\phi_0$ is not $\pi/2$.

Finally, the orthonormal basis functions on the truncated cone are:
\begin{equation}
f_{q, j}(\rho,\phi)=\frac{e^{iq \phi}}{\sqrt{2\pi}}\times\sqrt{\frac{2}{L}}\frac{\sin \frac{j \pi}{L}(\rho- \frac{R}{\sin \theta}+L)}{\sqrt{\sin \theta \rho}},
\end{equation}
where $q$ and $j$ are integer numbers. $f_{q, j}$ vanishes at the two edges of the truncated cone.
The energy spectrum on the cone geometry is obtained by expanding wave functions in terms of $f_{q, j}$.

\section{Analytic solution to the flat band of zero modes}
\label{appC}
In this appendix, we present analytic solutions to the flat band of zero modes in the nematic vortex of a polar $p$-wave superconductor.
Here we consider a disk geometry with no hole ($L=R$) and an infinite radius($R\rightarrow +\infty$).  The $q$th zero modes centered around the origin is presented in (\ref{dq}). The wave function $\psi_q(\rho)$ satisfies the following equation:
\begin{equation}
\big[\beta(\partial_\rho^2+\frac{1}{\rho}\partial_\rho-\frac{q^2}{\rho^2 })+\mu+\lambda_1\eta(\partial_\rho+\frac{1}{2\rho})\big]\psi_q(\rho)=0,
\label{psiq}
\end{equation}
which has the analytic solution
\begin{equation}
\psi_q(\rho)=\mathcal{N} J_q(\sqrt{1-(\lambda_1 \eta k_F/2\mu)^2}k_F\rho)\exp(-\frac{\lambda_1 \eta}{2\beta}\rho),
\end{equation}
where $J_q$ is the $q$th Bessel function. It is interesting to note that there is a zero-energy solution for every integer $q$ on the disk with infinite radius. As $|q|$ increases,  $\psi_q(\rho)$ moves away from the origin. On a finite-radius disk, there are also zero modes bound to the outer edge, which can hybridize with  $\psi_q(\rho)$ if the latter has a significant probability near the outer edge. Therefore, the number of zero modes on a finite-radius disk is finite, but it increases as $R$ increases. 

To gain a deeper insight, we revisit the problem of a superconducting vortex with an odd winding number in a polar $p$-wave superconductor\cite{Volovik1999}. The effective Hamiltonian is:
\begin{equation}
H_q=(\beta k^2-\mu)\tau_z+ [\frac{1}{2}(\kappa k_- + k_-\kappa)\tau_++h.c],
\end{equation}
where $\kappa=\eta \exp[i (2q+1) \phi]$ represents the vortex with winding number $2q+1$, where $q$ is an integer. In polar coordinates, we have:
\begin{equation}
\begin{aligned}
\frac{1}{2}(\kappa k_-+k_-\kappa)
=\eta e^{i 2q \phi}\big[-i (\partial_\rho + \frac{2q+1}{2\rho})-\frac{\partial_\phi}{\rho} \big].
\end{aligned}
\end{equation}
It is known that $H_q$ supports one zero energy mode\cite{Volovik1999}. We look for zero mode of the following form:
\begin{equation}
[ie^{i q \phi},e^{-i q \phi}]^{\text{T}}\psi_q(\rho).
\end{equation}
Here the wave function $\psi_q(\rho)$ should satisfy
\begin{equation}
\big[\beta(\partial_\rho^2+\frac{1}{\rho}\partial_\rho-\frac{q^2}{\rho^2 })+\mu+\eta(\partial_\rho+\frac{1}{2\rho})\big]\psi_q(\rho)=0,
\end{equation}
which shares exactly the same form as (\ref{psiq}). 
This provides another point of view on the reason why a polar $p$-wave superconductor supports a flat band of zero modes in the presence of a nematic vortex.

\bibliographystyle{apsrev4-1}
\bibliography{refs}

\end{document}